\documentclass[aps,pra,reprint]{revtex4-1}

\usepackage{amsmath}
\usepackage{amssymb}
\usepackage{amsfonts}
\usepackage{graphicx}
\usepackage{microtype}
\usepackage{geometry}
\usepackage[usenames]{xcolor}
\usepackage[textwidth=4cm,textsize=small,shadow]{todonotes}

\begin{document}

\title{Comment on ``Solvability of the two-photon Rabi
  Hamiltonian''}
\author{Andrzej J.~Maciejewski} \email{maciejka@astro.ia.uz.zgora.pl}
\affiliation{J.~Kepler Institute of Astronomy, University of Zielona
  G\'ora, Licealna 9, PL-65--417 Zielona G\'ora, Poland.}%
\author{Maria Przybylska}%

\email{M.Przybylska@if.uz.zgora.pl} \affiliation{ Institute
  of Physics, University of Zielona G\'ora, Licealna 9, 65--417
  Zielona G\'ora, Poland }%
\author{Tomasz Stachowiak} \email{stachowiak@cft.edu.pl}
\affiliation{%
  Center for Theoretical Physics PAS, Al. Lotnikow 32/46, 02-668
  Warsaw, Poland}

\begin{abstract}
  An implicit formula for the spectrum of the two-photon Rabi model
  was presented by Trav\v{e}nec in \cite{trawa} in analogy to the method
  of Braak \cite{braak}. The spectrum is given by common zeros of four
  functions $G_c(z, E)$ taken with a fixed value of $z\in\mathbb{C}$.
  We point out that all the $G_c(z,E)$ functions defined by Trav\v{e}nec
  are identically zero, thus the spectrum is not in fact given by their roots.
\end{abstract}

\maketitle

The problem at hand is characterised by a system of two differential
equations
\begin{equation}
  \label{tr}
  \begin{split}
    2g\phi''_1+\omega z\phi'_1+(2g z^2-E)\phi_1
    +\frac{\omega_0}{2}\phi_2 &= 0,\\
    2g\phi''_2-\omega z\phi'_2+(2g z^2+E)\phi_2
    -\frac{\omega_0}{2}\phi_1 &= 0.
  \end{split}
\end{equation}
The only singular point of the system is at infinity, and
it is irregular. This implies that all its solutions are given by
entire functions $\phi_1(z)$ and $\phi_2(z)$.  The problem is to find
those energy values $E=E_n$ for which the respective solution
$(\phi_1(z)$, $\phi_2(z))$ is such that its components $\phi_i(z)$ are
elements of Bargmann space $\mathcal{F}$ (see \cite{Bargmann:61::}). 
This requirement means additionally  that 
\[
\int_\mathbb{C}\overline{\phi_i(z)}\phi_i(z)\mathrm{e}^{-|z|^2}
\,\mathrm{d}\Re(z)\,\mathrm{d}\Im(z) <\infty, 
\]
for $i=1,2$.

To determine the spectrum four functions $G_{c}(z,E)$ were defined
in \cite{trawa}. The first of them $G_+(z,E)$ is defined by
\begin{equation}
  \label{eq:1}
  G_+(z,E):= \phi_2(\mathrm{i}z)-\phi_1(z),   
\end{equation}
where $(\phi_1(z),\phi_2(z))$ is the solution of system~\eqref{tr}
satisfying the following initial conditions
\begin{equation}
  \label{eq:2}
  \phi_1(0)=\phi_2(0)=1, \quad \phi_1'(0)=\phi_2'(0)=0. 
\end{equation}
From the above conditions and the symmetry of system \eqref{tr}
it follows that both functions $\phi_1$ and $\phi_2$ are even, so that
\begin{equation}
  \label{eq:4}
  \phi_1^{(2k+1)}(0)=\phi_2^{(2k+1)}(0)=0 
\end{equation}
for $k=0, 1, \ldots$.
According to \cite{trawa}, if $E$ belongs to the spectrum then
$G_+(z_0,E)=0$ for arbitrarily fixed $z_0\in\mathbb{C}$. We did not
find any justification for this claim. In \cite{trawa} it is only
remarked that the function $G_+(z,E)$ was constructed in full analogy
to that defined in \cite{braak} where the Rabi model was investigated.

We claim that function $G_{+}(z,E)$, as well as the remaining three
functions $G_c(z,E)$ defined in \cite{trawa}, vanish identically. This
is demonstrated below.

Eliminating $\phi_2$ from system~\eqref{tr} we find that
$f(z)=\phi_1(z)$ satisfies the following fourth order equation
\begin{equation}
  \label{iv}
  16g^2 f^{(\mathrm{iv})}+a_2(z)f''
  +a_1(z)f'+
  a_0(z)f=0,
\end{equation}  
where
\begin{align*}
    a_2(z)&= - 4(\omega^2-8g^2)z^2+16g\omega, \\
    a_1(z)&= 4z\left[16g^2-\omega(\omega-2E\right],\\
    a_0(z)&=16g^2(z^4+2)-16g\omega z^2-4E^2 +\omega_0^2.
\end{align*}
The crucial point is that function $G_+$, as well as the remaining
three functions satisfy the above equation. This can be checked
directly. This claim is true for an arbitrary choice of initial
conditions $\phi_i(0)$, $\phi_i'(0)$.

At this point it is important to observe that from \eqref{eq:2}
it follows that $G_+(z,E)$ satisfies the following conditions
\begin{equation}
  \label{eq:3}
  G_+(0,E)=G_+'(0,E)=G_+'''(0,E)=0.
\end{equation}
Moreover, taking initial condition \eqref{eq:2} and evaluating the
left hand sides of ~\eqref{tr} at $z=0$, we find that
\begin{equation}
  \label{eq:5}
  \phi_1''(0)= -\phi_2''(0) = \frac{2E-\omega_0}{4g}.  
\end{equation}
Thus $G_+''(0) = -\phi_2''(0)-\phi_1''(0)=0$. Hence, $G_+(z,E)=0$
identically because it is a unique solution of the linear homogeneous
equation~\eqref{iv} with vanishing initial condition. Using similar
reasoning we prove that remaining functions $G_c$ vanish identically.
Another simple way to check this fact is by direct computation from
the recurrence relations (13) in \cite{trawa}, and checking the series
expansions of $G_c$ around zero.  Expansions of $\phi_i$ up to $z^n$
will give all the terms of $G_c$ identically zero up to $z^n$. Hence,
taking limit $n\rightarrow \infty$ we obtain $G_c=0$,
regardless of the values of $E$ and the other parameters.

Let us remark that in \cite{braak} functions $G_c$ were introduced for
the study of the Rabi system, which has two regular singular points
and one irregular at infinity. In this system if a solution is entire
then it is automatically an element of the Bargmann space.
In \cite{braak} functions $G_c$ were used to glue together two local
solutions to obtain a global one which is entire,  whereas 
the solutions of \eqref{tr} are automatically entire. There is thus no basis
for deriving the spectrum from $G_c$ or its expansion.

\vspace{1em}
\section*{Acknowledgements}
 This
research has been supported by grant No.~DEC-2011/02/A/ST1/00208 of
National Science Centre of Poland.

\end{document}